\documentclass{PoS}

\usepackage{wrapfig}
\usepackage{url}

\title{LHC data challenges the contemporary parton-to-hadron fragmentation functions}

\ShortTitle{LHC data challenges the contemporary parton-to-hadron fragmentation functions}

\author{{David d'Enterria} \\
        CERN, PH Department, CH-1211 Geneva 23, Switzerland\\
        E-mail: \email{dde@cern.ch}}

\author{{Kari J. Eskola} \\
        Department of Physics, University of Jyv\"askyl\"a, P.O. Box 35, \\ FI-40014 University of Jyv\"askyl\"a, Finland \\ 
        Helsinki Institute of Physics, University of Helsinki, P.O. Box 64, FI-00014, Finland\\
        E-mail: \email{kari.eskola@jyu.fi}}

\author{{Ilkka Helenius} \\
        Department of Physics, University of Jyv\"askyl\"a, P.O. Box 35, \\ FI-40014 University of Jyv\"askyl\"a, Finland \\
        Helsinki Institute of Physics, University of Helsinki, P.O. Box 64, FI-00014, Finland\\
        E-mail: \email{ilkka.helenius@jyu.fi}}

\author{\speaker{Hannu Paukkunen} \\
        Department of Physics, University of Jyv\"askyl\"a, P.O. Box 35, \\ FI-40014 University of Jyv\"askyl\"a, Finland \\
        Helsinki Institute of Physics, University of Helsinki, P.O. Box 64, FI-00014, Finland\\
        E-mail: \email{hannu.paukkunen@jyu.fi}}

\abstract{We discuss the inclusive high-pT charged-particle production in proton-proton collisions 
at the LHC. The experimental data are compared to the NLO perturbative QCD calculations employing various sets 
of parton-to-hadron fragmentation functions. Most of the theoretical predictions are found to disastrously 
overpredict the measured cross sections, even if the scale variations and PDF errors are
accounted for. The problem appears to arise from the presently too hard gluon-to-hadron fragmentation functions.}

\FullConference{XXII. International Workshop on Deep-Inelastic Scattering and Related Subjects,\\
		28 April - 2 May 2014\\
		Warsaw, Poland}

\begin{document}

\section{Motivation}

The principal motivation for our study \cite{d'Enterria:2013vba} was the observation
that the next-to-leading order (NLO) perturbative QCD (pQCD) calculations for inclusive charged hadron production shown along with the
published p+p data from CMS \cite{CMS:2012aa} and ALICE \cite{Abelev:2013ala} appeared to clearly
overshoot the measurements at large transverse momentum ($p_T$). This called for a systematic study to chart the 
different sources of theory uncertainties and thereby, hopefully, identify the cause of the apparent mismatch.

\section{Framework}

\begin{wrapfigure}{r}{0.60\textwidth}
\centerline{\includegraphics[width=0.5\textwidth]{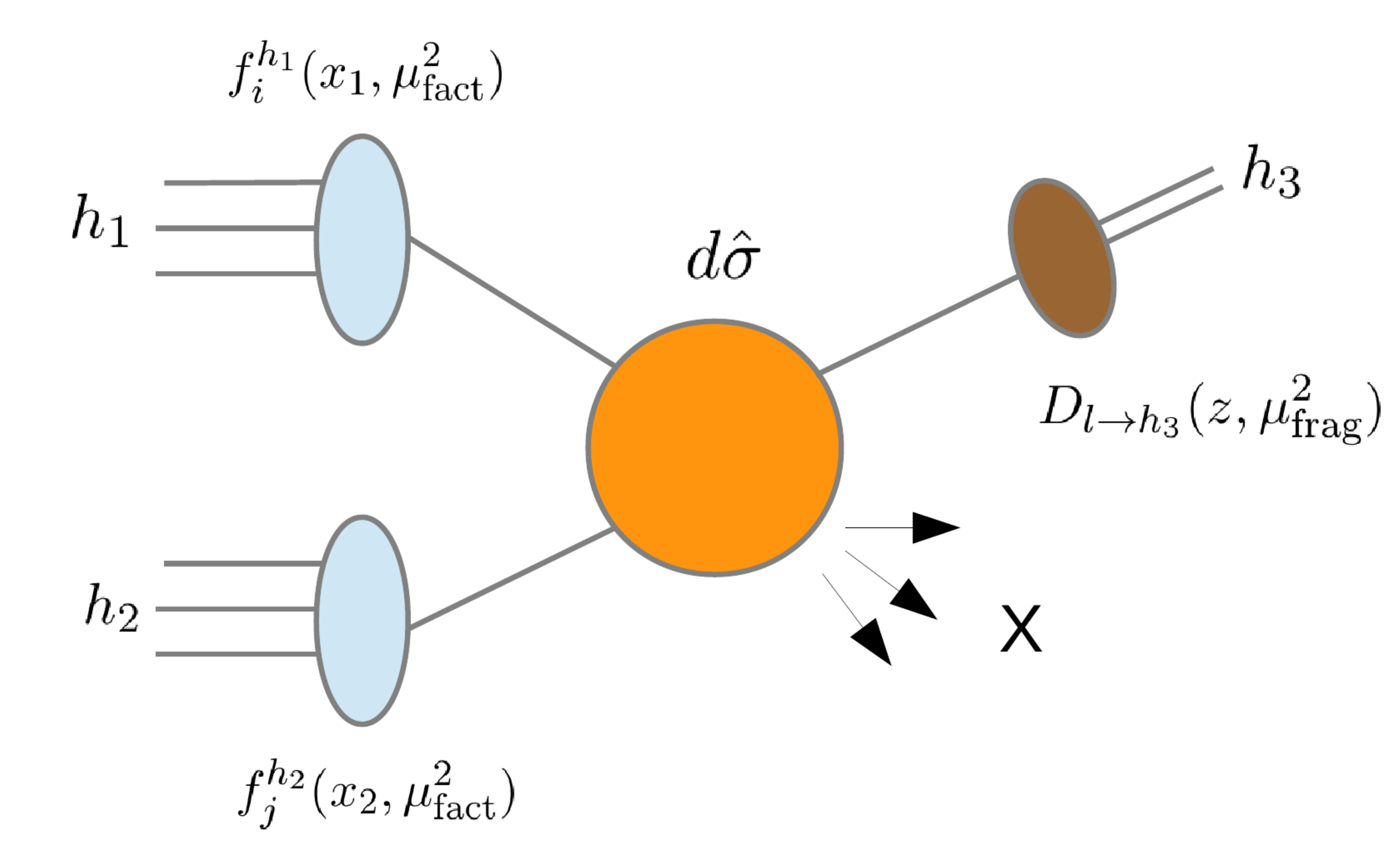}}
\caption[]{Ingredients of inclusive hadron production in hadronic collisions.}
\label{fig:schematic}
\end{wrapfigure}

\begin{wrapfigure}{r}{0.60\textwidth}
\centerline{\includegraphics[width=0.5\textwidth]{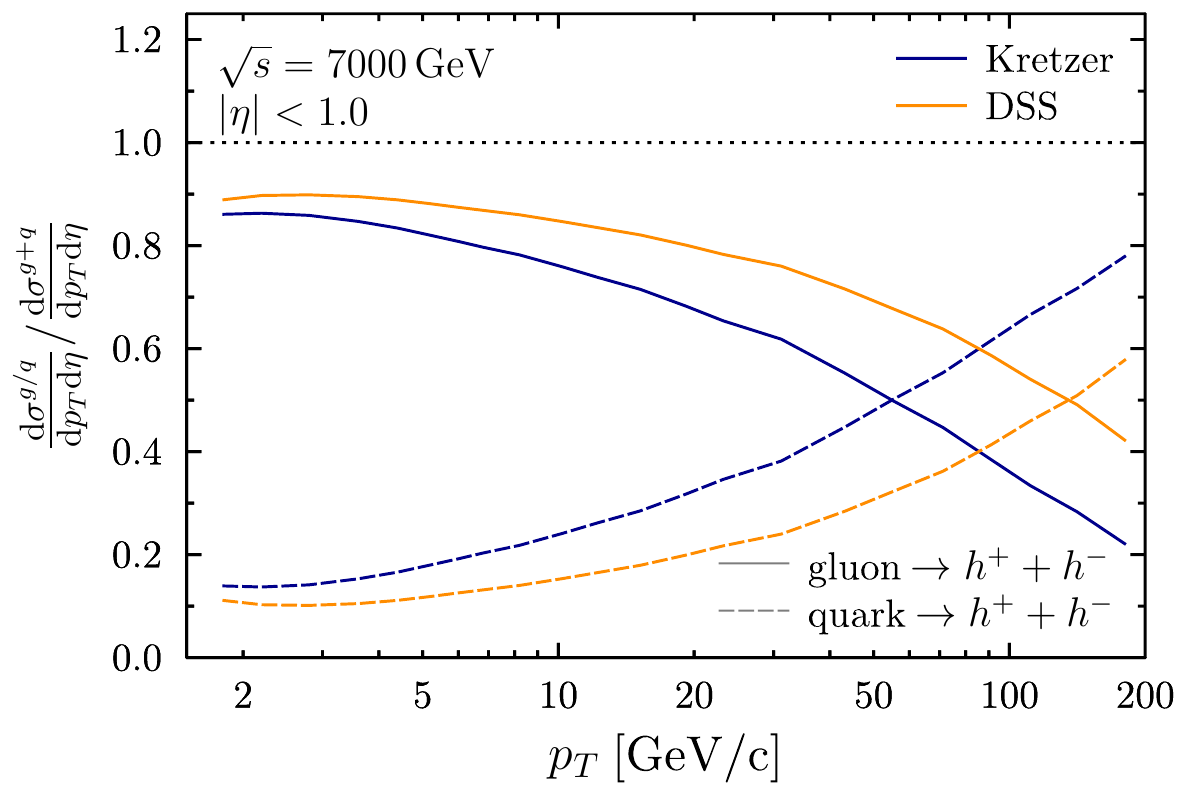}}
\caption[]{Relative importance of gluon (solid) and quark (dashed) fragmentation to the total
  charged-hadron yield with $\sqrt{s}=7000\,\rm{GeV}$ (right) at midrapidity. Calculation is
  shown for Kretzer (dark blue) and \textsc{dss} (orange) FFs. Figure from \cite{d'Enterria:2013vba}.}
\label{fig:ff_comparison3}
\end{wrapfigure}

The inclusive high-$p_T$ charged-particle ($h_3$) production in collision of two hadrons $h_1$ and $h_2$ can be computed as
a convolution of the initial-state parton distribution functions $f_i(x,\mu^2_{\rm fact})$ (PDFs), final-state parton-to-hadron fragmentation
functions $D_{k \rightarrow h_3}(z,\mu^2_{\rm frag})$ (FFs), and partonic coefficient functions $d\sigma^{i+j \rightarrow k + X}(\mu^2_{\rm ren},\mu^2_{\rm fact},\mu^2_{\rm frag})$
as
\begin{equation}
d\sigma(h_1+h_2 \rightarrow h_3 + X) = \sum_{ijk} f_i^{h_1} \otimes d\sigma^{i+j \rightarrow k + X} \otimes f_j^{h_2} \otimes D_{k \rightarrow h_3},
\end{equation}
where $\mu^2_{\rm ren},\mu^2_{\rm fact}$, and $\mu^2_{\rm frag}$ denote the renormalization, factorization,
and fragmentation scales, respectively. Figure~\ref{fig:schematic} presents a schematic illustration of the process ingredients.
%\begin{figure}[tbhp]
%\centering
%\includegraphics[width=0.49\textwidth]{output.pdf}
%\caption{} 
%\label{fig:schematic}
%\end{figure}
In practice, we evaluate the cross sections to NLO in strong coupling $\alpha_s({\mu^2_{\rm ren}})$
utilizing the public \textsc{incnlo} code \cite{INCNLO,Aversa:1988vb}. For the PDFs we
use \textsc{ct10nlo} \cite{Lai:2010vv} and its error sets. The FF uncertainty is estimated
by performing the calculations with various parametrizations: Kretzer (\textsc{kre})~\cite{Kretzer:2000yf}, \textsc{kkp}~\cite{Kniehl:2000fe},
\textsc{bfgw}~\cite{Bourhis:2000gs}, \textsc{hkns}~\cite{Hirai:2007cx}, \textsc{akk05}~\cite{Albino:2005me}
\textsc{dss}~\cite{deFlorian:2007aj,deFlorian:2007hc}, and \textsc{akk08}~\cite{Albino:2008fy}. From all these,
only \textsc{hkns} offers a possibility to estimate the propagation of FF uncertainties to further observables.
\begin{figure}[tbhp]
\centering
\includegraphics[width=0.49\textwidth]{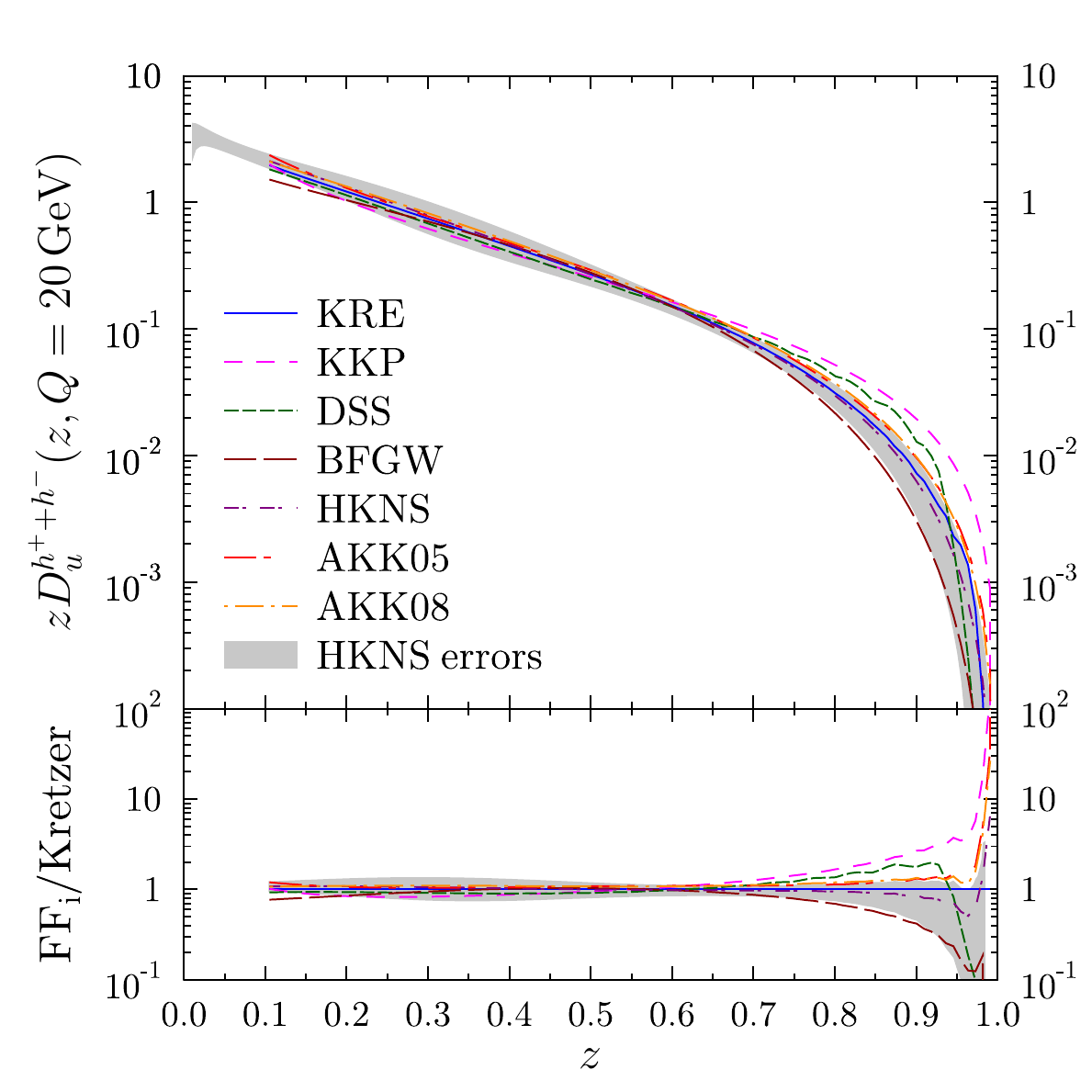}
\includegraphics[width=0.49\textwidth]{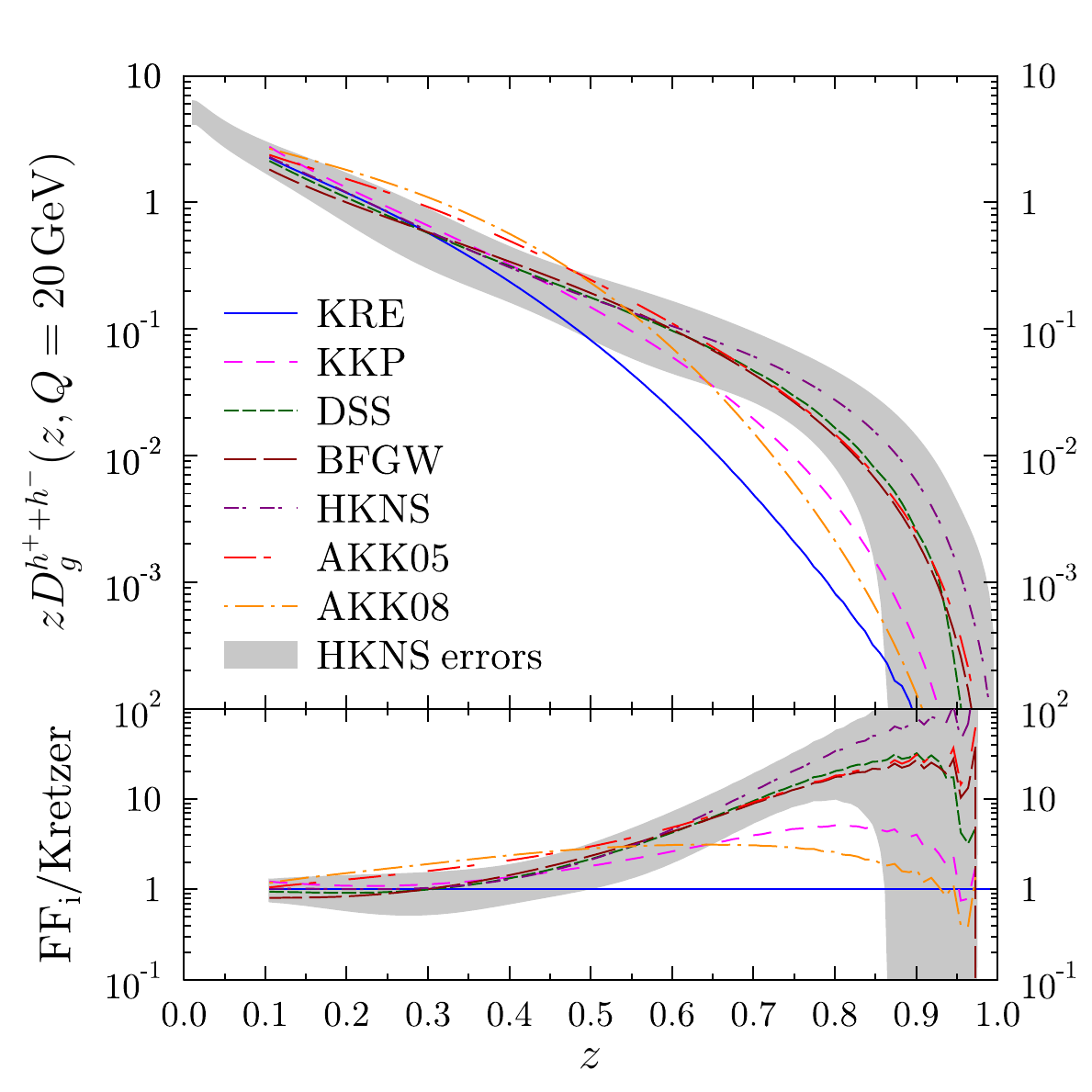}
\caption{The absolute charged-hadron FFs for $u$-quarks (left) and gluons (right) at $\mu_{\rm frag}=20\,\rm{GeV}$, and their ratios
to Kretzer FFs. Figure from \cite{d'Enterria:2013vba}.}
\label{fig:ff_comparison}
\end{figure}
We illustrate the differences in these FFs by plotting, in Figure~\ref{fig:ff_comparison}, the FFs for $u$-quarks and
gluons. While the $u$-quark FFs are all rather tightly packed together, the spread among the different gluon FFs is huge.
The main reason for the large differences is that the $e^+e^-$ annihilation data which constitutes the bulk of
the FFs constraints are predominantly sensitive to the quark fragmentation, leaving the gluons rather unconstrained.
To improve the situation \textsc{dss} and \textsc{akk08} included also hadroproduction data from RHIC, SPS and Tevatron,
but predominantly at rather low values of $p_T$ where the NLO pQCD calculations are in doubt (see later).
Along with the LHC p+p runs at several center-of-mass energies it has now become possible to fit the FFs at higher $p_T$ and
also cross-check between different experiments. As illustrated in Figure~\ref{fig:ff_comparison3}, the
gluon fragmentation dominates the cross sections up to the highest measurable values of $p_T$ and the large spread of
gluon FFs seen in Figure~\ref{fig:ff_comparison} should therefore make a difference at the LHC.

Our default choice for the involved QCD scales is $\mu_{\rm ren}=\mu_{\rm fact}=\mu_{\rm frag}=p_T$, and
we explore the sensitivity of the NLO calculation to this particular choice by varying the scales independently
by a factor of two. However, we exclude the variations with
$$
\frac{\mu_{\rm ren}}{\mu_{\rm frag, fact}} = 4 \,\, {\rm or} \,\, \frac{1}{4},
$$
as this gives rise to artificially large logarithms
$
\log ({\mu^2_{\rm ren}}/{\mu^2_{\rm fact}}) %\,\, \log ({\mu^2_{\rm ren}}/{\mu^2_{\rm frag}}),
$
and
$
\log ({\mu^2_{\rm ren}}/{\mu^2_{\rm frag}}),
$
in terms involving partonic splitting functions $P_{ij}$. Schematically,
{\footnotesize
\begin{equation}
\frac{\alpha_s(\mu_{\rm ren}^2)}{2\pi} \log \left( \frac{\hat s}{\mu^2_{\rm fact,frag}} \right) P_{ij}
\approx
\frac{\alpha_s(\mu_{\rm fact,frag}^2)}{2\pi} \log \left( \frac{\hat s}{\mu^2_{\rm fact,frag}} \right) P_{ij} 
+
\frac{\alpha_s^2(\mu_{\rm fact,frag}^2)}{2\pi} \frac{\beta_0}{4\pi} \log \left( \frac{\mu_{\rm fact,frag}^2}{\mu_{\rm ren}^2} \right)
\log \left( \frac{\hat s}{\mu^2_{\rm fact,frag}} \right) P_{ij},
\end{equation}
}
which follows from the QCD renormalization group equation ($\beta_0$ is the first term
in the QCD $\beta$-function and $\hat s$ denotes the partonic center-of-mass energy).
Whereas the first term tends to moderate effect of varying $\mu_{\rm fact,frag}$, the second
term can become large.

\newpage
\section{Results}

\begin{figure}[tbhp]
\centering
\includegraphics[width=0.75\textwidth]{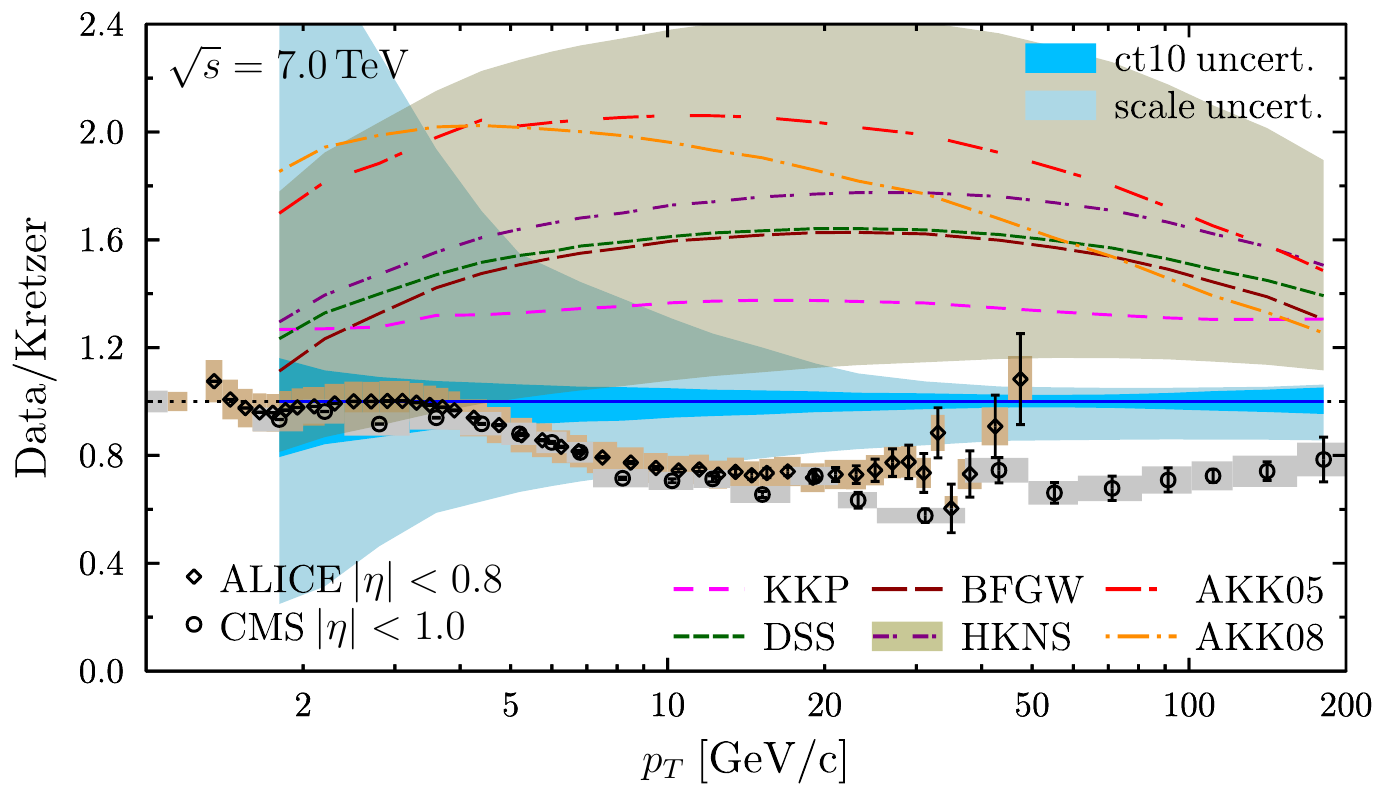}
\caption{Comparison between the CMS and ALICE $7 \, {\rm TeV}$ data and the calculations
with various FFs. Figure adapted from \cite{d'Enterria:2013vba}.} 
\label{fig:7TeV}
\end{figure}

Figure~\ref{fig:7TeV} shows a representative example of what we find when contrasting the NLO
calculations against the measurements. The data as well as all the calculations with various FFs have
been normalized by the results obtained with the Kretzer FFs. The main points we want to emphasize are:
\begin{itemize}
 \item The CMS and ALICE data are in a fair agreement.
 \item The PDF-originating uncertainty (dark blue band) turns out rather small, around 5\% at high $p_T$.
 \item The scale uncertainty (light blue band) is huge at small $p_T$, but levels off beyond $p_T \sim 10 \, {\rm GeV}$.
 \item The calculation with Kretzer FFs gives the best description of the data, being around 20\% above the data beyond $p_T \sim 10 \, {\rm GeV}$.
 The data-to-theory ratio is remarkably flat despite the fact that the absolute cross section drops around eleven 
 orders of magnitude.
 \item All other FFs overshoot the data even more. The error band of \textsc{hkns} is large but not large enough
 to enclose the data.
\end{itemize}
Although we presented here only the $7\, {\rm TeV}$ LHC data as an example, the situation remains more or less the 
same when considering the other LHC energies and also with the data from the CDF collaboration \cite{Aaltonen:2009ne}
(see Ref.~\cite{d'Enterria:2013vba} for more details). It should be noted that related processes like inclusive jet \cite{Chatrchyan:2012bja,Aad:2011fc}
or isolated photon production \cite{Chatrchyan:2011ue,Aad:2011tw} at the LHC are in agreement with the NLO calculations.
This reinforces the idea that the mismatch between the LHC data and the NLO calculations is indeed due to the
current gluon FFs which have either (almost) no data constraints or have been constrained at very low $p_T$ where the
scale uncertainty is enormous. The scale uncertainty is, however, not the only reason why the region below $p_T \sim 10 \, {\rm GeV}$
should be discarded from charged-hadron FF fits: The qualitative difference between the $(K^++K^-)/(\pi^++\pi^-)$ and $(p^++p^-)/(\pi^++\pi^-)$
ratios measured by ALICE \cite{Abelev:2014laa,RobertoPreghenellafortheALICE:2013yua} indicate that the baryon production
at this low-$p_T$ region cannot be considered being just independent parton-to-hadron fragmentation but different (collective?)
physics seems to be involved. From $p_T \sim 10 \, {\rm GeV}$ onwards such differences appear to disappear and the NLO calculations, like the ones presented here,
should be adequate.

\begin{figure}[tbhp]
\centering
\includegraphics[width=0.49\textwidth]{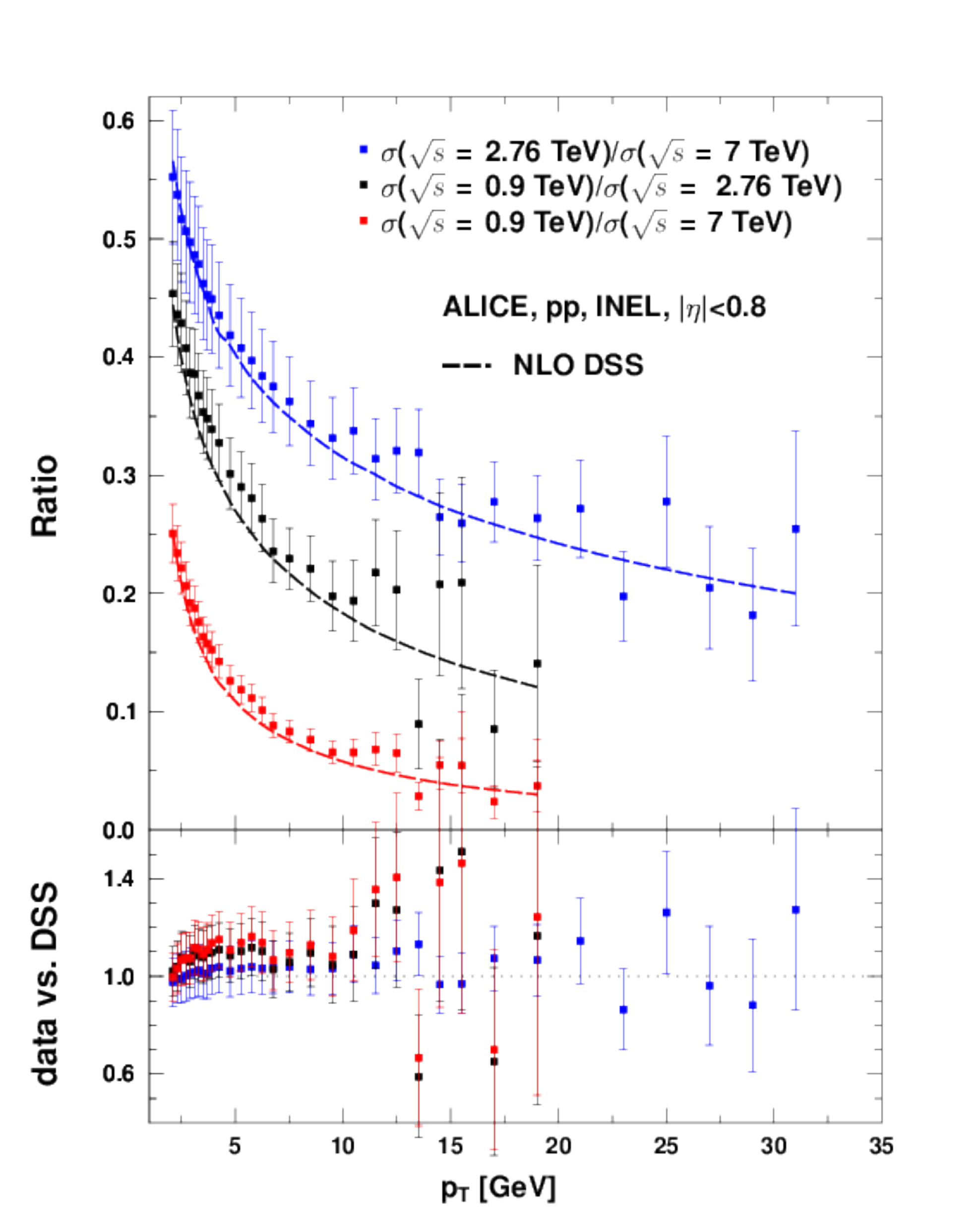}
\includegraphics[width=0.49\textwidth]{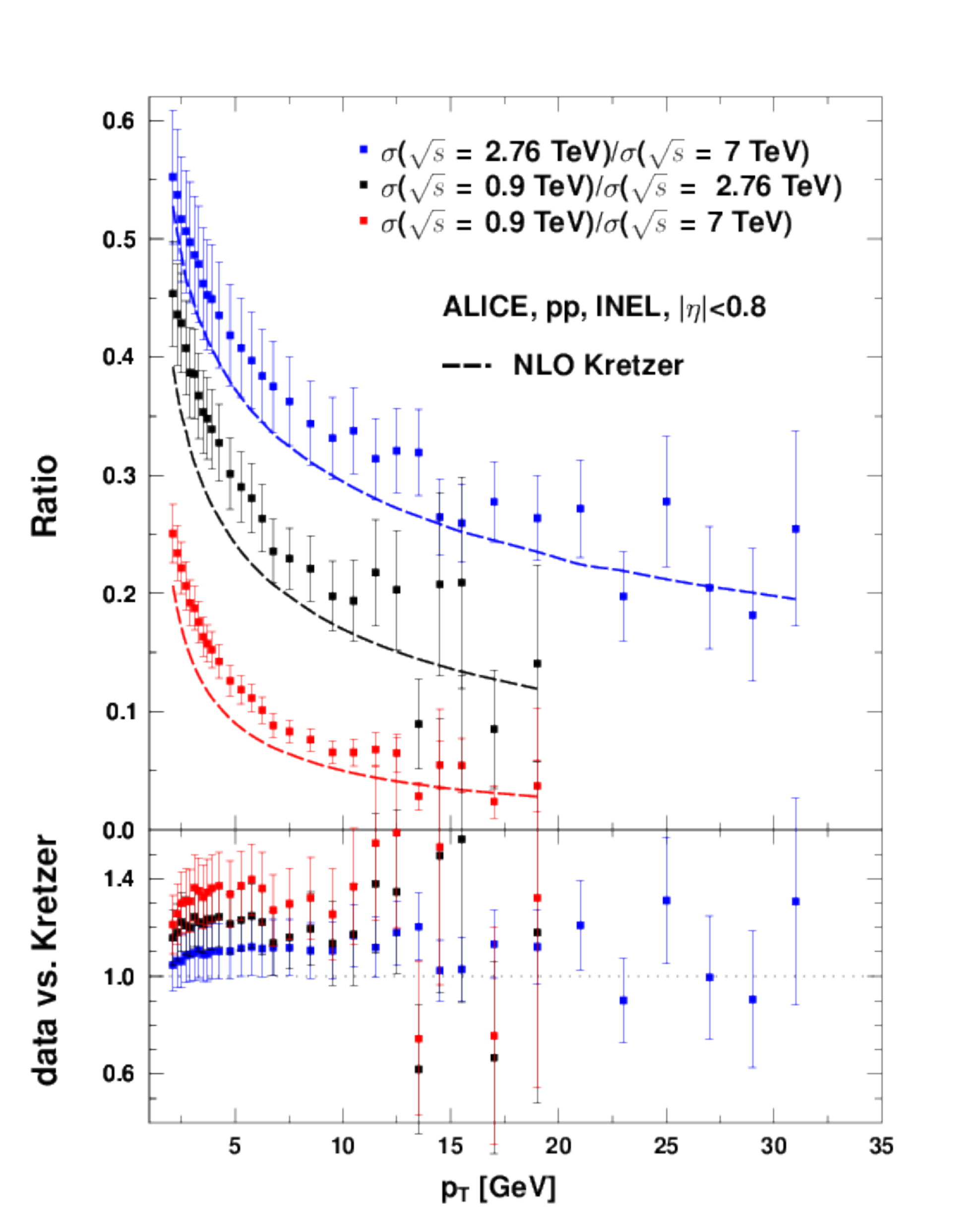}
\caption{Ratios between the charged-hadron yields at different center-of-mass energies. The
data points are from ALICE \cite{Abelev:2013ala} (constructed by dividing the cross sections
and adding the uncertainties in quadrature) and the calculations (dashed lines) are obtained
using the \textsc{dss} (left) and Kretzer (right) FFs.} 
\label{fig:ff_comparison5}
\end{figure}

Despite the significant data-to-theory mismatch in the case of absolute cross sections
it has been noticed \cite{Abelev:2013ala} that the ratios of cross section between different
$\sqrt{s}$ are, however, much better described by the NLO pQCD.	This is demonstrated
in Figure~\ref{fig:ff_comparison5} presenting some ALICE data and calculations with two different FFs, Kretzer and \textsc{dss}.
Indeed, even \textsc{dss} which grossly overshoots the absolute spectra (see Figure~\ref{fig:7TeV}) is, more or less, consistent
with the data. This follows from the fact that the cross-section ratios in Figure~\ref{fig:ff_comparison5} are more sensitive to the
shape of the gluon FFs and not that much to their absolute magnitude. 

\section{Summary}

In conclusion, we have found that none of the current sets of FFs can optimally describe the 
LHC (or Tevatron) data for inclusive charged hadron production at $p_T \gtrsim 10 \, {\rm GeV}$.
Below $p_T \sim 10 \, {\rm GeV}$ the scale uncertainty is enormous prohibiting
to make practically any conclusion and, in addition, in this low-$p_T$ region there are 
evidence for excess baryons which do not seem to originate from independent
parton-to-hadron fragmentation. For these reasons we conclude that only the
region $p_T \gtrsim 10 \, {\rm GeV}$ should be used in the forthcoming fits of
charged-hadron FFs.

\section*{Acknowledgments}
\noindent We acknowledge the financial support from the Magnus Ehrnrooth Foundation (I.H.) and from the
Academy of Finland, Project No. 133005.

\end{document}